# A liquid metal encapsulation for analyzing porous nanomaterials by atom probe tomography

Se-Ho Kim[1,†], Ayman A. El-Zoka[1,†,*], Baptiste Gault[1,2,*]

[1]Max-Planck-Institut für Eisenforschung, Düsseldorf, Germany.

[2]Department of Materials, Royal School of Mines, Imperial College London, London, UK.

[†]These authors contributed equally to this work as co-first authors.

[*]Corresponding author. Email: a.elzoka@mpie.de (A.A.E.-Z.); b.gault@mpie.de (B.G.)

## Abstract

Analyzing porous (nano)materials by the atom probe tomography has been notoriously difficult. The electrostatic pressure intensifies stress at voids which results in premature failure of the specimen, and the electrostatic field distribution near voids lead to aberrations that are difficult to predict. Here we propose a new encapsulating method for a porous sample using a low-melting-point Bi-In-Sn alloy, known as Field's metal. As a model porous sample, we used single-crystalline wüstite following direct hydrogen-reduced into iron. The complete encapsulation is performed using in-situ heating on the stage of the scanning-electron microscope up to approx. 70˚C. No visible corrosion nor dissolution of the sample occurred. Subsequently specimens are shaped by focused ion beam milling under cryogenic conditions at -190˚C. The proposed approach is versatile, can be applied to provide good quality atom probe datasets from microporous materials.

# Introduction

Atom probe tomography (APT) (Devaraj et al., 2018) provides the 3D compositional and, sometimes, crystallographic information (Gault et al., 2012) from a material's microstructure with sub-nanometer resolution (De Geuser & Gault, 2020; Jenkins et al., 2020). To enable field evaporation, the material of interest must be shaped into a sharp needle with a uniform geometry, and a radius of curvature in the range of 100 nm at the tip. However, not all materials of interest are fully consolidated or solid. A lot of recent studies have sought to apply APT on porous materials (El-Zoka et al., 2017), nanoparticles (Lim et al., 2020), and materials with cracks (Meisnar et al., 2015) or voids (Wang et al., 2020). Leaving these pores and cracks open would introduce stress concentrators that make specimen preparation difficult, and complicate the reconstruction of the analyzed specimens due to trajectory aberrations and premature fractures during analysis (Pfeiffer et al., 2015). It has been established in the field of APT to deal with this problem by effectively eliminating these pores/voids/cracks with foreign chemically yet distinguishable materials using electrodeposition for instance (Ayman A. El-Zoka et al., 2018; S.-H. Kim et al., 2020), electron beam induced deposition (Barroo et al., 2020), and other physical vapor deposition methods (Felfer et al., 2014).

Although the surface of sputter-coated samples appears dense, this process cannot penetrate and efficiently fill voids within porous materials. The chemical vapor deposition from C-containing materials typically leads to non-homogeneous field-evaporation conditions and highly-complex mass spectra due to carbon-containing molecular ions (Prosa et al., 2010; Nishikawa et al., 2006). The key challenge in designing a pore infiltration approach, is devising a method that does not alter the chemistry and/or structure of the original material, while ensuring maximum pore infiltration. This could be difficult for metals that tend to corrode or dissolve due to their relative reactivity, such as iron. The electrodeposition methods reported previously, with e.g. Cu and Ni from sulfate solutions work perfectly for noble metal materials such as Au (Ayman A. El-Zoka et al., 2018), Pt (A. A. El-Zoka et al., 2018), and Pd (Kim et al., 2018). However, with a reactive metal such as iron, electrodeposition of Cu and Ni will more likely result in dissolution of Fe due to galvanic replacement (Kim et al., 2018), or lead to the formation of FeS on the surfaces during the negative polarization of the sample (Lyon, 2010).

Herein, we demonstrate herein the use of BiSnIn, a low melting point alloy known as Field's metal, for infiltrating porous Fe samples formed by direct hydrogen-gas reduction of an iron oxide single crystal. We drew inspiration from a previous report that re-deposition of a fusible alloy enabled embedding of freestanding nanoparticles (Kim et al., 2019), and devised a suitable infiltration method for porous transition metals. We use a heating stage inside the scanning-electron microscope (SEM) to heat up the Field's metal and fill micron-sized pores inside of the FeO-sample. Then, a cryogenically-cooled stage is used during the focused ion beam (FIB) milling of needle-shaped specimen suitable for atom probe from the Fe-Field metal composite. Cryo-FIB has recently proven advantageous for avoiding ingress of Ga (Dantas De Morais et al., 2014) and modifications of the specimen's composition and structure during the preparation of APT specimens (Rivas et al., 2020; Lilensten & Gault, 2020; Chang et al., 2019) and for the preparation of specimens from low-melting point materials (Schreiber et al., 2018; El-Zoka et al., 2020). By using this approach, we successfully analyzed the microporous, reduced iron that exhibits a bimodal pore distribution, and show data both the Fe and the filling material. APT analysis enables the identification of structure and chemistry of the hydrogen-reduced Fe samples (Se-Ho Kim et al., 2020), including oxide inclusions with a clear advantage when it comes to the identification of peaks in the mass spectra. Our method highlights the overlooked beneficial properties of low melting point metals in APT analysis of high surface area samples used in various technological field.

## Materials and methods

A size of 5 x 5 x 0.5 mm$^3$ single crystalline FeO (wüstite orientation (100), MaTech) was sliced into small pieces and 20 g of the as-received Bi-In-Sn fusible alloy (Field's metal, Alfa Aesar) was cut into a 0.5 x 1.0 x 0.5 cm$^3$ in size. The as-received wüstite was reduced in a furnace with pure dihydrogen gas with a flow rate of 30 L hr$^{-1}$ at 700 °C for 2 hrs. After the complete reaction, the reduced iron piece was left in the furnace to cool down to room temperature and collected.

For APT specimen preparation, we used a SEM-FIB FEI helios nanolab 600, equipped with a stage that can perform heating and be cryogenically-cooled by a cold gaseous nitrogen (Gatan C1001 system). Specimens from the wüstite before reduction were prepared using the lift-out protocol outlined in Ref. (Thompson et al., 2007). For the porous material, the fusible alloy, the porous material, and the Cu clip were loaded in this same microscope. All scanning electron micrographs were taken at 30 kV and 1.6 nA. A final annular ion-beam milling process at 30 kV and 48 pA was performed. A Ga-contamination cleaning protocol at 5 kV was not done on the fabricated APT specimens since a cryo-prepared sample has less critical issue with Ga implementation due to the decrease of the Ga penetration rate (Lilensten & Gault, 2020; Thompson et al., 2007).

Following heating to room-temperature in the SEM-FIB, the APT specimens were taken out under ambient conditions and loaded inside a local electrode atom probe (LEAP) 5000 XS system (CAMECA). A specimen base temperature of 50 K, a pulse frequency of 100 kHz, a detection rate of 0.5 %, and a pulse energy of 10 pJ were set for APT measurement. The Atom Probe Suite 6 developed by CAMECA was used for the 3D reconstructions and data analysis.

## Results

### *Before and after H reduction on a single crystalline wüstite*

Figure 1a shows the morphology of the as-received wüstite sample. Following preparation by in-situ lift-out, the APT measurement was successful (Figure 1b and 1c). The 3D atom map of a dataset from the wüstite is shown in Figure 1d. As expected, Fe and O are homogenously distributed, and the atomic composition of Fe and O is 50.6 and 49.4 at. %, respectively.

After the H reduction process of the wüstite, the iron become a sponge-like structure with large cracks and micron-sized open pores (see Figure 1e–g) (Matthew et al., 1990). In our preliminary trials, the standard lift-out method is no longer applicable to fabricate needle-shaped APT specimens from the reduced iron, since there are too many defects with a critical size within the specimen's failure. The specimen itself cannot stand its weight during the Ga-ion annular milling process and mechanically fails. When a specimen survived the preparation, voids and cracks cannot tolerate the field evaporation and eventually result early fracture of the specimen in the atom probe.

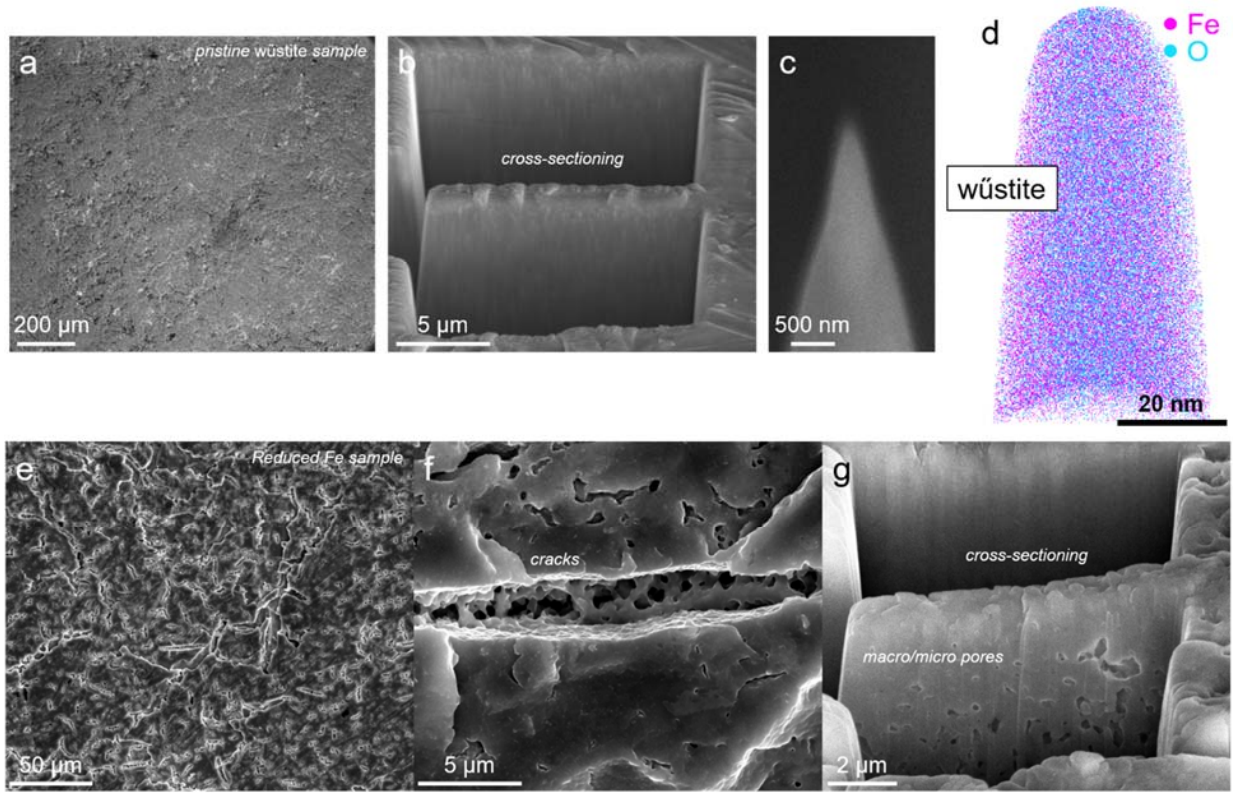

**Figure 1.** Characterization on wüstite before and after H-reduction. (a) Surface and (b) cross-sectional SEM images of as-received wüstite. (c) A final APT specimen of wüstite. (d) 3D atom map of wüstite from the acquired dataset. (e,f) Surface and (g) cross-sectional SEM images of as-reduced iron.

*Encapsulation of porous sample with Bi-In-Sn alloy and APT specimen preparation*
First, the sliced fusible alloy is placed on a blank Si holder and tilted to 52º towards an e-beam column in the FIB to mill a clear pattern from an uncontrollable milling behavior of a fusible alloy. Square trenches (15 x 3 µm²) are milled 10 µm in depths on the front and the back sides of the interested region using a Ga-ion beam current of 2.8 nA at 30 kV. Because of ion-beam induced heating locally and re-deposition, an uncontrollable milling behavior appears and therefore a vicinity of the trenches is uneven (Figure 2a). The stage is then tiled it back to 0º and the L-shape horizontal cut is made at the bottom of the lamella (Figure 2b). The sample is lifted out in-situ using the micro-manipulator, and welding with Pt/C assisted by the Ga ion beam (Figure 2c).

In order to prevent the spreading of the liquid-metal droplet on a Cu clip during heating, a trench (2 x 15 x 1 µm³) is milled as shown in Figure 2d. The lifted-out lamella is carefully attached on the side of the trench using FIB-Pt/C deposition (Figure 2e–g). Then a lamella is lifted out of the reduced porous iron sample and the lamella is placed on top of the fusible alloy (Figure 2h–i).

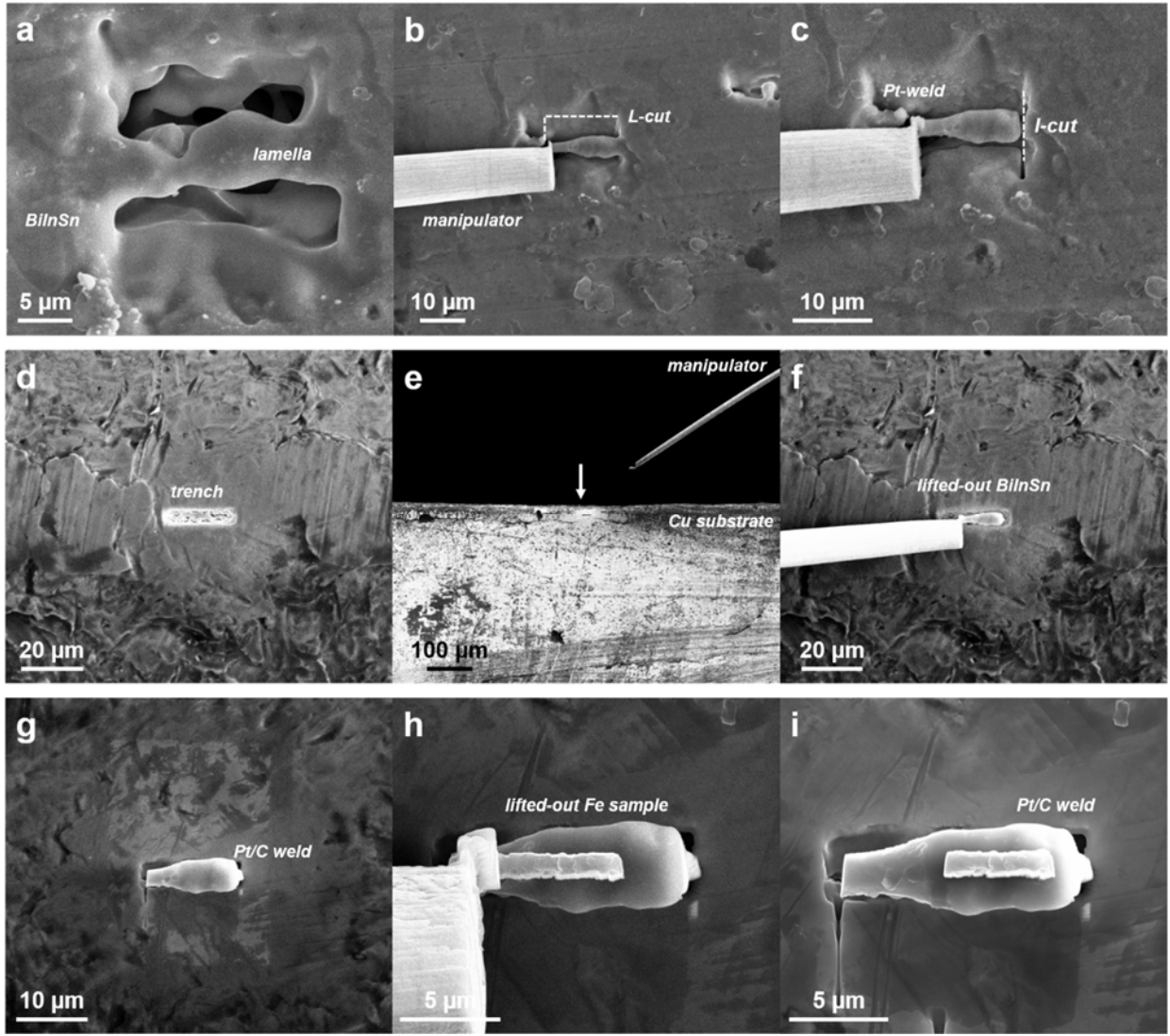

**Figure 2.** A lifted-out Bi-In-Sn alloy and porous iron on a Cu clip before in-situ heating and cooling process. (a) The Bi-In-Sn alloy before lift-out. (b) L-horizontal cut is made and a micro-manipulator is inserted to lift the lamella. (c) I-vertical cut releases the lamella from the bulk. (d) A prior trench on the Cu clip. (e) A manipulator with the sample attached is inserted (f-g) The lamella is mounted on a Cu trench. (h) The porous Fe lamella is mounted on the top. (i) the stacked sample before heating process.

After the attachment, the Cu clip is taken out and the FIB stage is switched to the heating/cryo stage. Using the temperature controller, the stage is heated up to melt the fusible alloy which has the melting point of 62 °C. The lamella of the alloy gradually melts and subsequently becomes a droplet-like feature at 70 °C (see Figure 3a–c). We let the liquid sufficiently infiltrate the porous iron for approx. five minutes, and then the stage is cooled back down to room temperature.

Figure 3d and 3e show that the cooling below the melting point of the fusible alloy led to the solidification of the melt and the formation of a composite structure. The open micro pores are filled with the Field's metal matrix. Figure 3f shows the cross-sectioning of the porous iron embedded in the fusible alloy matrix. The cross-sectional composite lamella is prepared with a sufficiently a low ion-beam current (93 pA) to avoid potential melting. No open pore or crack are observed in the SEM, and the back-scattered electron (BSE) signals showed no difference between the iron and the matrix (not shown here) indicating that the melted alloy filled the voids.

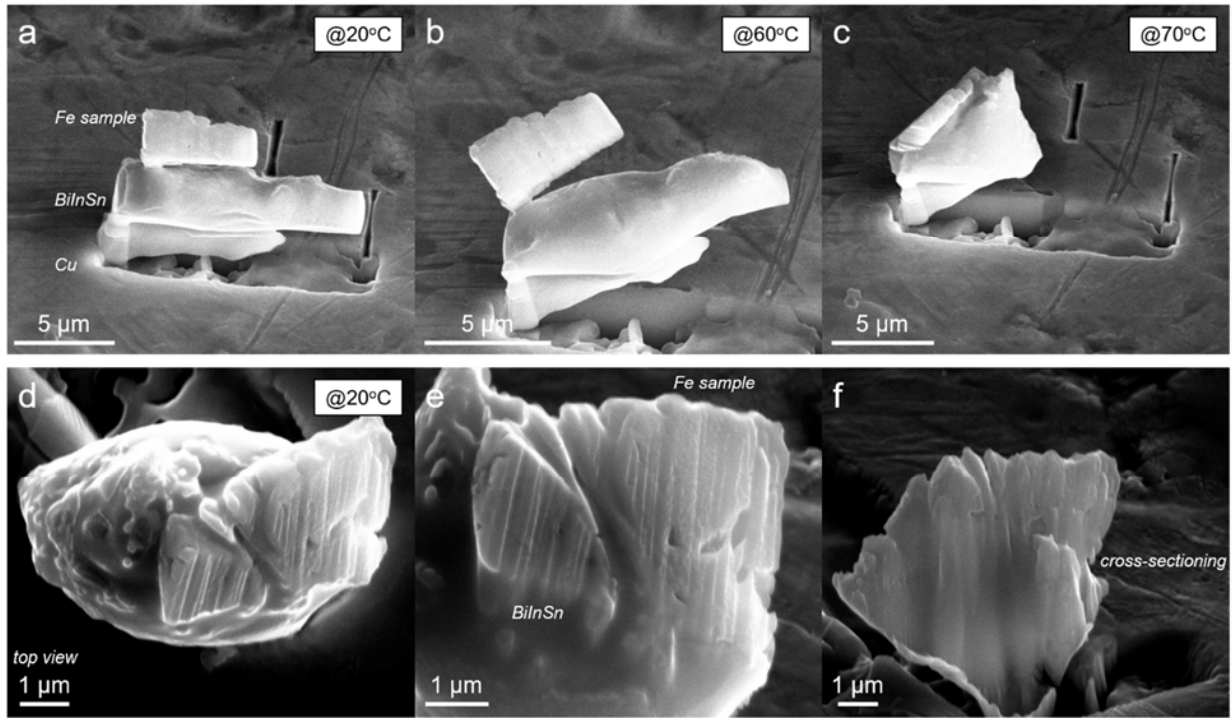

**Figure 3.** In-situ experiment: (a-c) heating up to 70 °C and (d-f) cooling back to 20 °C.

Kim et al. reported uncontrollable Ga-ion milling behavior when a diameter size of the fusible alloy is less than 1 µm due to the size-dependent reduction of the melting temperature (Kim et al., 2019); therefore, after the composite sample is mounted on a commercial Si micro-tip coupon, the stage is cooled to cryogenic temperature, at -190 °C, to avoid the melting of the Field's metal within the needles (Figure 4a). Figure 4b and 4c show the controlled ion milling, which is performed at 30 kV and 48 pA. Overall, with this approach, a porous sample difficult to analyse by APT was fabricated into a dense needle-shaped specimen suitable for APT.

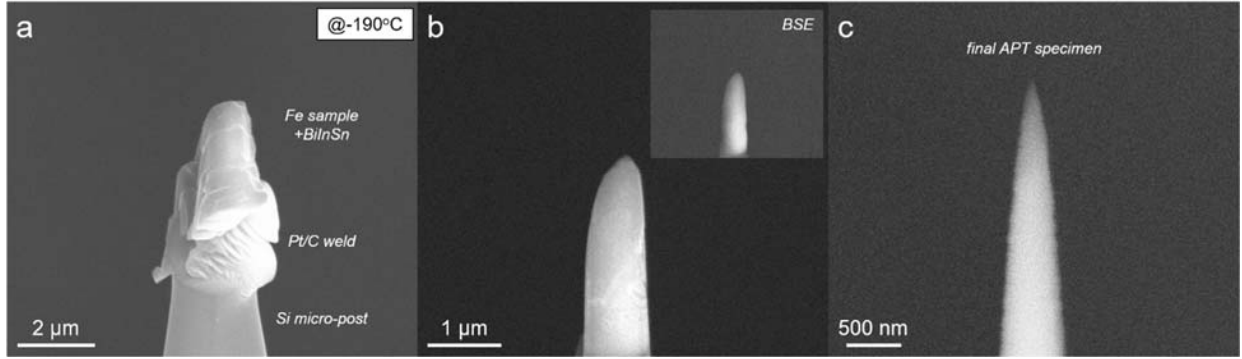

**Figure 4.** In-situ cryogenic annular milling of the composite. (a) The specimen post is cooled down to cryogenic temperature. (b) After a rough annular FIB milling step. Inset shows BSE image of the specimen. (c) A final APT specimen milled at cryogenic temperature.

*APT analysis*

Figure 5a shows the 3D atom map acquired from the reduced iron. The atomic compositions of Fe and O are to be 99.87 and 0.13 at. %, indicating that most of O was removed by the hydrogen reduction process. From the 2D contour map of Fe calculated within the region-of-interest shown in a cyon box, a strong pole signal indicates that the reduced sample has a body centered cubic structure that corresponds to α-Fe. No constituent elements of the fusible alloy are detected which implies that a liquid metal penetration does not proceed into the iron.

Within the α-Fe, interestingly an oxide particle is detected as shown in Figure 5b. Although the as-received sample was high-purity, it still contained impurities such as Mg and Ca which are commonly known as gangue elements in an iron ore (Se-Ho Kim et al., 2020). In the α-Fe, these impurities are in a precipitate-like feature with a diameter of approximately 4 nm. During the H-reduction process, gangue oxide cannot be reduced and impurities that cannot be expelled from the metal at the reduction front are trapped inside the α-Fe, forming small particles.

It is reported that the Bi-In-Sn alloy has three distinctive phases namely $BiIn_2$, $In_2Sn_9$, and $In_3Sn$. Here, in Figure 5c, the 3D atom map of the fusible alloy shows a homogeneous elemental distribution and the composition is 91.92, 8.06 and 0.02 at.% for Sn, In, and Bi, respectively, which corresponds to the $In_2Sn_9$ phase. Within the matrix, no iron-related peak nor impurities oxide peaks from the porous iron are detected indicating that no galvanic replacement happened during the sample preparation.

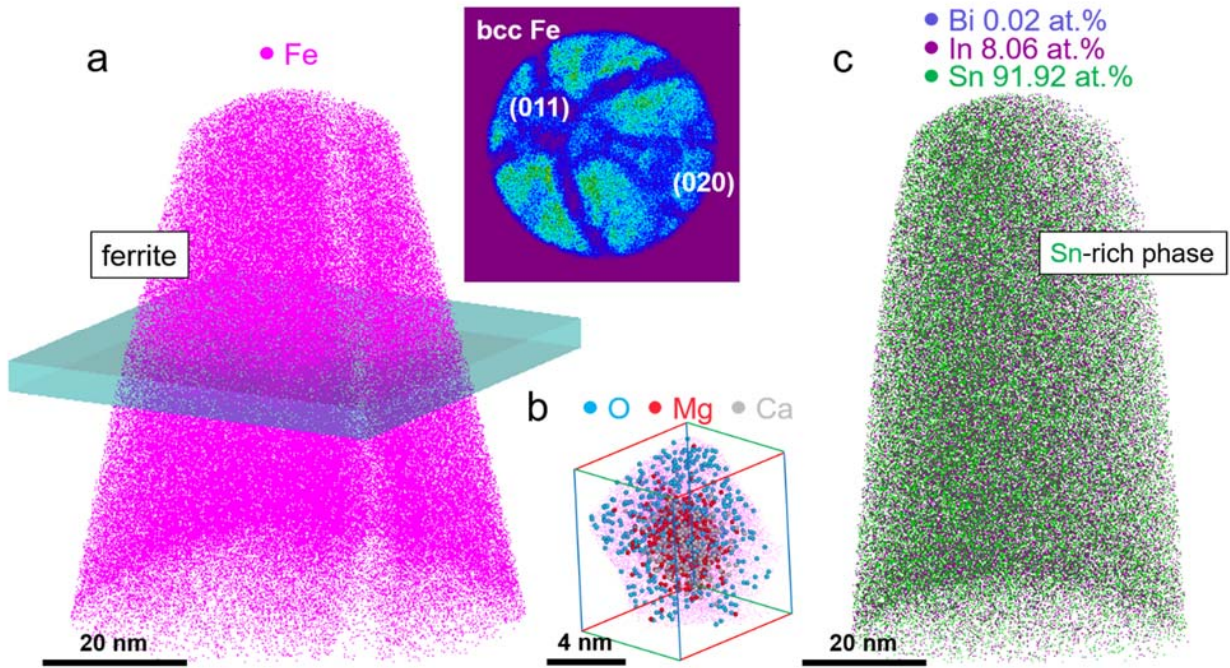

**Figure 5.** (a) 3D atom map of the α-Fe sample. Pink dots represent reconstructed Fe atoms. Inset shows 2D desorption map histogram of Fe with crystallographic pole features. (b) Extracted atom map of impurity oxide region. Blue, red, and gray dots represent reconstructed O, Mg, and Ca atoms. (c) 3D atom map of the Bi-In-Sn alloy. Blue-gray, purple, and green represent reconstructed Bi, In, and Sn atoms.

## Summary and discussion

To summarize, the overall process for filling voids in the materials with the fusible alloy is schematically shown in Figure 6. First, a trench was milled on a commercial Cu stub (CAMECA) to load the lifted-out fusible alloy and subsequently the porous material. The lift-out process was performed according to Ref.(Miller et al., 2007; Thompson et al., 2007). The stage temperature first raised up to 70 ºC in order to melt the fusible alloy. After the stacked samples are mixed into a single composite, the stage is cooled down to room temperature. After solidification of the molten metal, a region of interest was extracted using the micromanipulator and attached to a support using ion-beam induced Pt/C deposition. Once three sliced samples are attached onto the support, the stage was cooled down to cryogenic temperature (approx. -190 ºC) to avoid ion-beam induced melting of the Field's metal.

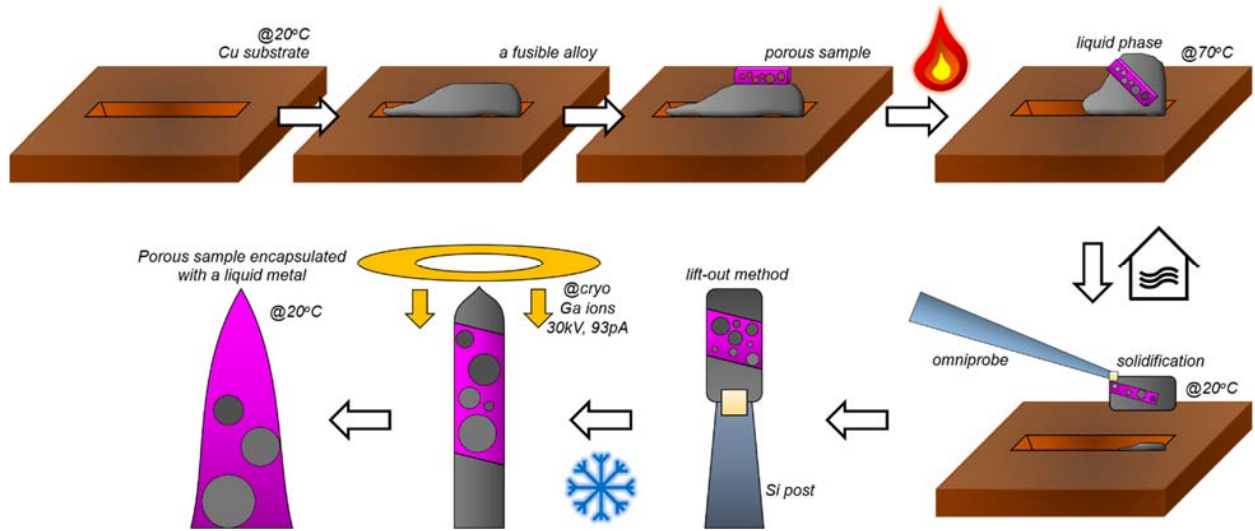

**Figure 6.** A schematic illustration of encapsulation of porous sample in a fusible alloy.

The combination of in-situ FIB heating and cooling, combined with a liquid metal is a versatile approach to infiltrate and encapsulate micro-to-nanoporous materials and enable the APT measurement successful. An advantage of this method is that the encapsulating material is conductive. According to Seol et al., the surface modification from the conductive layer coating improves the overall evaporation sequences (Seol et al., 2016). Unlike ceramic coating approaches (Sundell et al., 2019; Webel et al., 2021), a high quality data can be acquired because a fusible alloy has relatively high thermal (0.19 W/cm·°C) and electrical conductivity (0.57 nΩ·m) which allows to dissipate the energy and electron quickly to the base (Lipchitz et al., 2015). This can explain the high (3:3) success rate, with over 10 million ions collected for each specimen, and without premature or micro-fractures. A difficulty with this protocol is to position the material of interest towards the specimen's apex, otherwise only the matrix is collected during the atom probe measurement.

Development of other alloys for encapsulation could be very beneficial to further optimize the preparation. One of the main criteria to be taken into consideration is that none of the metals in the alloy are miscible with the material to be analyzed. This was the case for Fe and BiSnIn (von Goldbeck & von Goldbeck, 1982a, 1982b, 1982c). The melting point of BiSnIn is also suitable in our case so as not to coarsen the porous structure. We imagine that a dip-coating process can be performed for aggregated powder in the future using a suitable liquid metal. After lifting the interested lamella, the sample can be immersed in a preheated liquid metal and it allows penetrates pores or spaces inside (see Figure 7).

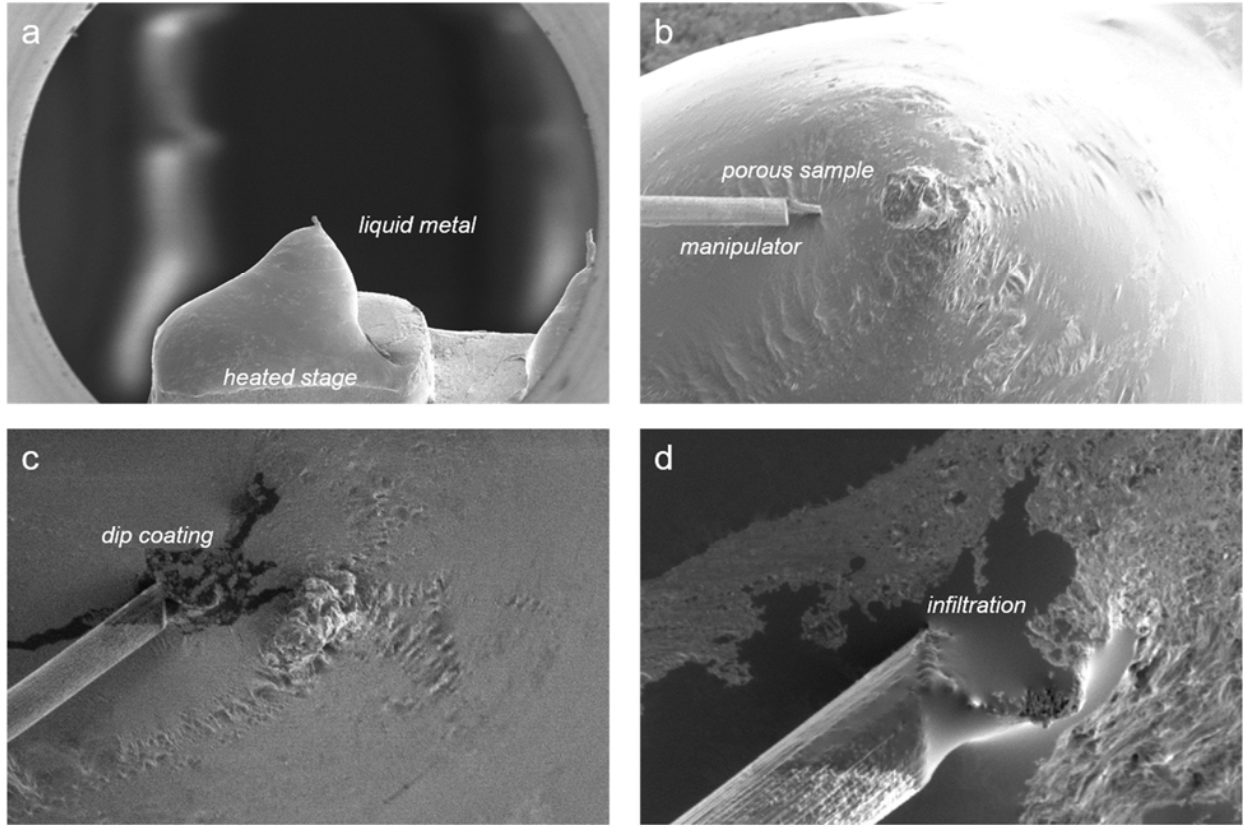

**Figure 7.** A modified encapsulating approach for a porous nanomaterial using a liquid metal.

# Conclusion

We have demonstrated a new method for preparing APT specimens from a sample containing micro- to nanopores. A sponge-like iron sample prepared by hydrogen reduction from a single-crystal of wüstite was infiltrated and embedded in a fusible Bi-In-Sn alloy. The complete encapsulation is achieved by melting and solidifying the alloy using an in-situ SEM-FIB. No corrosion or dissolution of the sample is observed during the sample preparation and high-quality APT dataset is obtained. We expect the demonstrated method allows to explore variety of porous materials in APT.

# Acknowledgement

The authors are grateful to Uwe Tezins, Andreas Sturm and Christian Broß for their support to the APT and FIB facilities. We also thank Dr. Xue Zhang, Dr. Michael Rohwerder and Dr. Dierk Raabe and for providing samples and collaborating on the direct hydrogen reduction of iron ore project. S.-H.K., A.E.-Z., and B.G. acknowledge financial support from the ERC-CoG-SHINE-771602.